# *Current-voltage characteristics of layered cuprate superconductors with intrinsic Josephson junctions*

**T.B. Charikova, V.N. Neverov, M.R. Popov, S.D. Popov, N.G. Shelushinina**

M.N. Mikheev Institute of Metal Physics, Ural Branch, Russian Academy of Sciences, 18, S. Kovalevskoy St., Ekaterinburg, 620077, Russia

**Abstract**

A brief review of the results for the current-voltage characteristics study in highly anisotropic cuprate superconductors is presented, considering the existence of the intrinsic Josephson effect in multilayer HTSC compounds. We analyzed the observations of multi-branch current-voltage characteristics for current in the *c*-direction both for the hole-doped and electron-doped cuprate SCs which conveniently viewed as a stack of intrinsic Josephson junctions, where the $CuO_2$ layers serve as superconducting plates and the buffer layers as dielectric interlayers. It was shown that the tunneling characteristics of intrinsic Josephson junctions in optimally doped p- and n-type cuprates indicate a predominantly d-wave symmetry of the order parameter, demonstrating high conductivity in the subgap regime: $U < 2\Delta/e$. The analysis carried out confirms the theoretically predicted power-law dependences $U \sim I^{\gamma}$ with $\gamma = (0.33 - 0.50)$ in the hole-doped Bi- and Tl-compounds, which is typical for SCs with monotonic $d_{x^2-y^2}$-wave symmetry of the order parameter and with $\gamma = (0.70 - 0.76)$ for electron-doped SCs with a non-monotonic d-wave symmetry within the framework of the two-band/two-gap model.

## 1. Introduction

Oxide superconductors are layered compounds with building blocks consisting of conductive $CuO_2$ layers separated by buffer layers that serve as charge reservoirs (see monographs [1–4] for a detailed description). Highly anisotropic high-temperature superconductors (HTSCs) can be considered as a "package" of superconducting $CuO_2$ layers coupled by Josephson interactions [5–8]. The new properties of these materials compared to single Josephson junctions are associated with their multilayer structure and the atomic thickness of the superconducting layers.

The importance of the mechanism of interlayer coupling (in the *c*-direction) was especially emphasized already at the dawn of the study of cuprate HTSCs, when Anderson developed the model of interlayer tunneling [9]. This model considers tunneling processes along the *c*- axis both in the superconducting (tunneling of Cooper pairs) and in the normal (single-particle tunneling) states of a layered superconductor. Since the transfer between layers occurs through successive tunneling of charge carriers, the corresponding multilayer structures can be considered as a set of Josephson tunneling junctions. These junctions are called the intrinsic Josephson junctions [10,11].

The intrinsic Josephson effect, conditioned by the tunneling of charge carriers in the superconducting (SC) state of cuprate multilayer HTSCs, has been intensively studied in recent decades. The distinctive structure of the current-voltage characteristics for current in the *c*-direction, with a large (up to several hundred) number of hysteresis resistive branches, is most pronounced in the highly anisotropic Bi- and Tl-systems [12,13] (see also reviews [10,11] and references there). The fact that internal tunneling of Cooper pairs does take place has been experimentally confirmed for many cuprate systems: for $Bi_2Sr_2CaCu_2O_8$ (BSCCO) [10–15], $Tl_2Ba_2Ca_2Cu_3O_{10}$ (TBCCO) [12], $Bi_2(Sr_{2-x}La_x)CuO_6$ (BSLCO) [16], $La_{2-x}Sr_xCuO_4$ (LSCO) [17], $La_{1.6-x}Nd_{0.4}Sr_xCuO_4$ (LNSCO) [18], for electron-doped cuprates $Pr_{2-x}Ce_xCuO_4$ (PCCO) [19] and $Sm_{2-x}Ce_xCuO_4$ (SCCO) [20], as well as for the magnetic superconductor $RuSr_2GdCu_2O_8$ (RSGCO) [21].

The further structure of this review is as follows. After a brief presentation of the resistively and capacitively shunted junction (RCSJ) model for a superconductor-insulator-superconductor (SIS) Josephson junction (Section 2), experimental results on multi-branch current-voltage (CV) characteristics for both hole-doped and electron-doped SC cuprates are presented, including our results for NCCO films (Section 3).

In Section 4, the existing theoretical concepts for the coherent and noncoherent tunneling processes in SIS Josephson junctions with *d*-wave symmetry of the order parameter are presented, and experimental results for the scaling behavior of the subgap conduction at intrinsic CV characteristics in the layered cuprate compounds of *p*- and *n*-types are analyzed. Section 5 summarizes the results of the review.

## 2. RCSJ model of Josephson junction (theory)

The first, stationary (dc) Josephson effect [22] is as follows: through a tunnel contact between two superconductors (a contact of superconductors through a thin layer of insulator, SIS junction), in the absence of an externally applied potential difference, an undamped (nondissipative) superconducting current (supercurrent) can flow:

$$I_s = I_c \, sin\Delta\varphi \tag{1}$$

Here $I_c$ is the critical current of Josephson junction (the maximum value of supercurrent that a Josephson junction can support), $\Delta\varphi$ is the phase difference of the order parameter (of the Ginsburg–Landau wave function) in the two sides of the junction (with the normal state resistance $R$):

$$I_c = \frac{\pi\Delta_0^2}{4eT_cR}, \tag{2}$$

where $\Delta_0$ is the superconducting gap at $T = 0$, $T_c$ is the critical temperature.

In this case, we are talking specifically about the properties of the junction, and the current $I_c$ is much (by orders of magnitude) less than the destruction current of Cooper pairs in a bulk superconductor (Ginsburg-Landau critical current $I_{GL}$): $I_c \ll I_{GL}$.

The term *Josephson effect* currently refers to a set of phenomena that occur in contacts of two superconductors through a weak links (see, for example, the review by Likharev [23], as well as the books by Tinkham [3] or Schmidt [24]).

As long as the external current $I$ is less than the critical $I_c$, the entire current is superconducting, $I = I_s$. If $I > I_c$, then in addition to the superconducting current, a normal current component, $In$, associated with dissipation, and a voltage $V$ at the contact, appear.

Within the resistive model (see, for example, [24]) for a SIS junction, the Josephson junction can be represented as a parallel connection of an ideal (non-dissipative) contact and the normal resistance $R$.

In the general case, it is necessary also to take into account the capacitive effects, since the very design of the Josephson junction, in which the two sides are separated by a dielectric layer, resembles a capacitor. Although direct current is not possible through a capacitor, an alternating bias current $I_d = CdV/dt$ (with $C$ being a capacitance) can flow through it due to recharging of the capacitor plates. A complete description is usually provided by the model of resistively and capacitively shunted junction (RCSJ) [3,24] (see Fig. 1a).

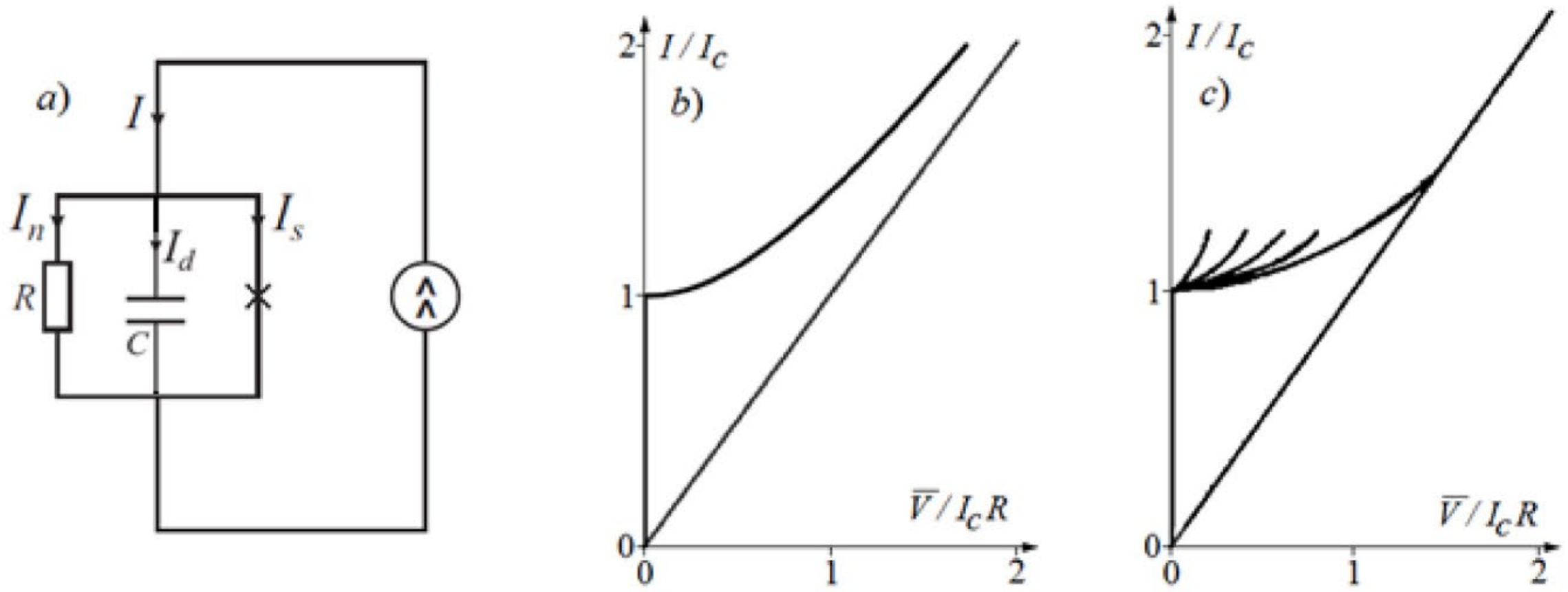

**Fig. 1.** (a) - RCSJ model of a Josephson junction connected to a circuit with a current source. The Josephson junction is indicated by a cross (from Ref. [24]). (b) - Current-voltage characteristic of a single Josephson junction. The voltage at the contact occurs at $I > Ic$ (from Ref. [24]). (c)- Generalization to the case of the intrinsic Josephson effect in multilayer system (scheme).

The total current $I$ through the system is now the sum of the superconducting current $I_s = I_c \sin\Delta\varphi$, the normal dissipative current $I_n = V/R$, where $R$ is the normal state resistance, and the bias current through the capacitor $I_d$:

$$I = I_s + I_n + I_d. \tag{3}$$

The presence of capacitance in the Josephson contact leads to an ambiguous form of its current voltage characteristic: the dependence $V(I)$ becomes different for the cases of increasing and decreasing $I$, i.e. hysteresis occurs. Quantitatively, the hysteresis properties are determined by the value of the McCumber parameter [25,26]:

$$\beta_c = \left(\frac{2e}{\hbar}\right) I_c C R^2, \tag{4}$$

which is a measure of the damping of the Josephson junction.

All of the above applies to a single Josephson junction. The CV characteristics of arrays of Josephson tunnel junctions in multilayer HTSCs consist of several branches, each corresponding to one, two, three, etc. individual junctions switching to quasiparticle state as the external current exceeds the corresponding critical current (see reviews [10,11] and references therein). This situation is shown schematically in Fig. 1c.

The dynamic behavior of layered structures with Josephson connections has been analyzed in detail, for example, in Refs. [14,27] based on a model with multiple sequential Josephson junctions. It is shown that the conditions under which multiple resistive branches are observed on the current-voltage characteristic are associated both with the independent dynamic behavior of each junction and with the presence of hysteresis (with the value of McCumber parameter $\beta_c$).

## 3. Experimental results on current-voltage characteristics

### *3.1 Intrinsic Josephson junction current-voltage characteristics of hole-doped cuprates*

Experimental works on the effects of carrier tunneling along the *c*-axis in highly anisotropic HTSCs clearly showed that the materials behave as a set of SIS Josephson junctions (see references in the Introduction). Adjacent superconducting layers in a high-temperature superconductor are weakly coupled by the Josephon effect, causing the single crystals to act essentially as the vertical stacks of hundreds of intrinsic Josephon junctions.

Typical CV characteristics for current flow in the direction of the *c*-axis of layered HTSC systems (see reviews [10,11]) demonstrate a series of sequential Josephson junctions: they have many resistive branches and a clearly defined hysteresis. In this section, we are discussing some results obtained for stacks of intrinsic Josephson junctions in cuprate p-type superconductors. The characteristic structure of the current-voltage characteristics for current in the direction of the *c*-axis, with a large number of hysteresis resistive branches is clearly manifested in the most anisotropic layered Bi- and Tl-systems.

Fig. 2 shows the first three branches of the CV characteristic of an intrinsic junction $Tl_2Ba_2Ca_2Cu_3O_{10+\delta}$ step stack at $T$ = 4.2 K. Measurements of the complete current-voltage characteristic for the sample with a stepped structure demonstrate the existence of 130 branches in the resistive state.

After exceeding the critical current of an individual junction in the stack, the CV characteristic exhibits a voltage jump $U_c$ which is about $2\Delta/e$, the value expected for a standard SIS tunnel junction. The voltage jump is approximately 27 mV in TBCCO step stacks which corresponds to roughly 50% of $2\Delta/e$.

Note that quasiparticle branches of the intrinsic junctions show a gradual current rise in the subgap regions. If the junctions in the stack are essentially equal and independent, then at fixed bias current the voltage in the *n*th branch should simply be *n* times the voltage of the first branch, at least in the first few branches.

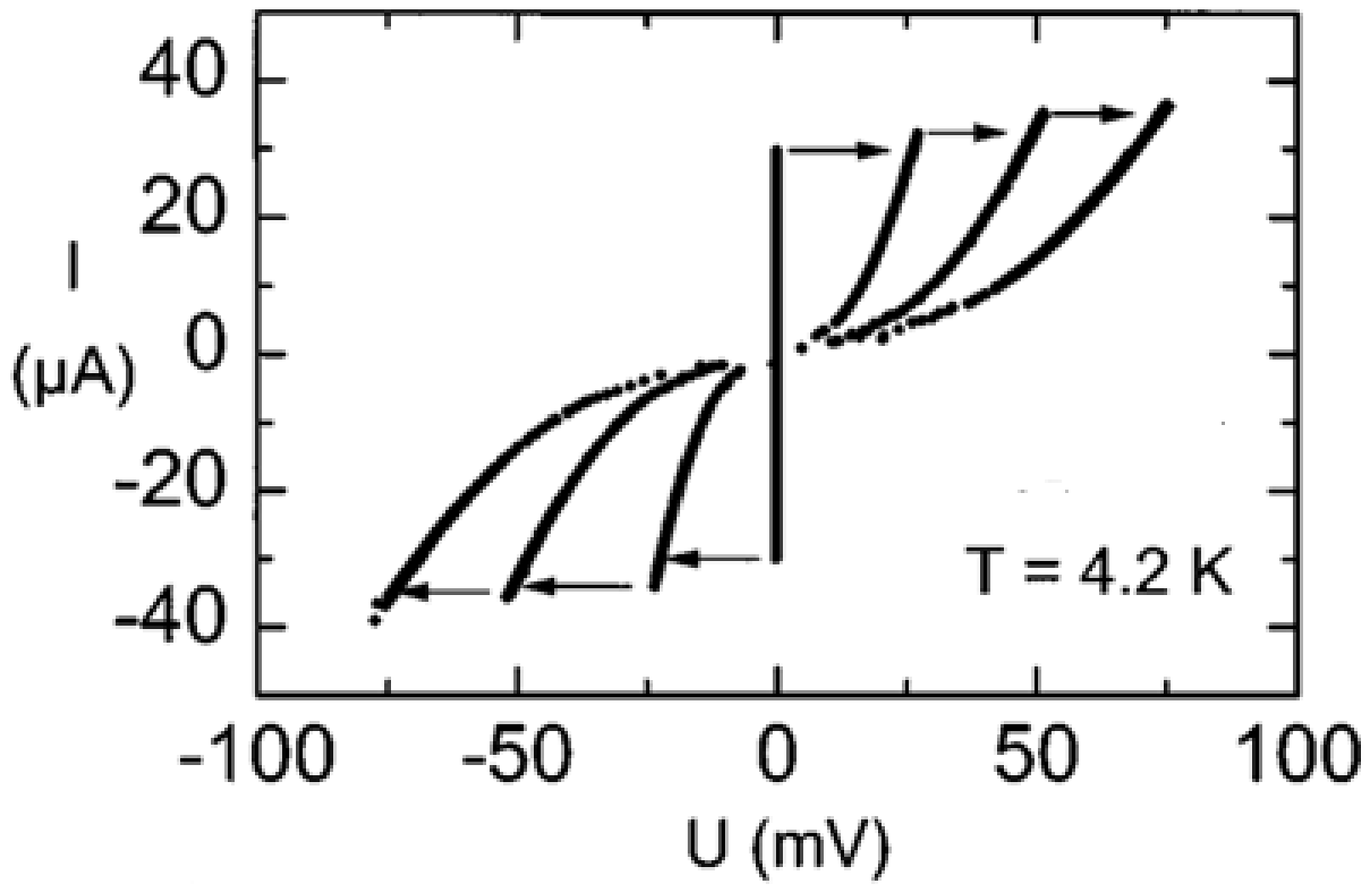

Fig. 2. First three branches of the CV characteristic of a $Tl_2Ba_2Ca_2Cu_3O_{10+\delta}$ step stack with a total number of 130 junctions. Arrows indicate voltage switching (after [10]).

Fig. 3 shows data from the work of Heim et al. [11] on the intrinsic Josephson effect in the $Bi_2Sr_2CaCu_2O_8$ system. Properly prepared mesa structures on BSCCO single crystals act as stacks of intrinsic Josephson tunnel junctions.

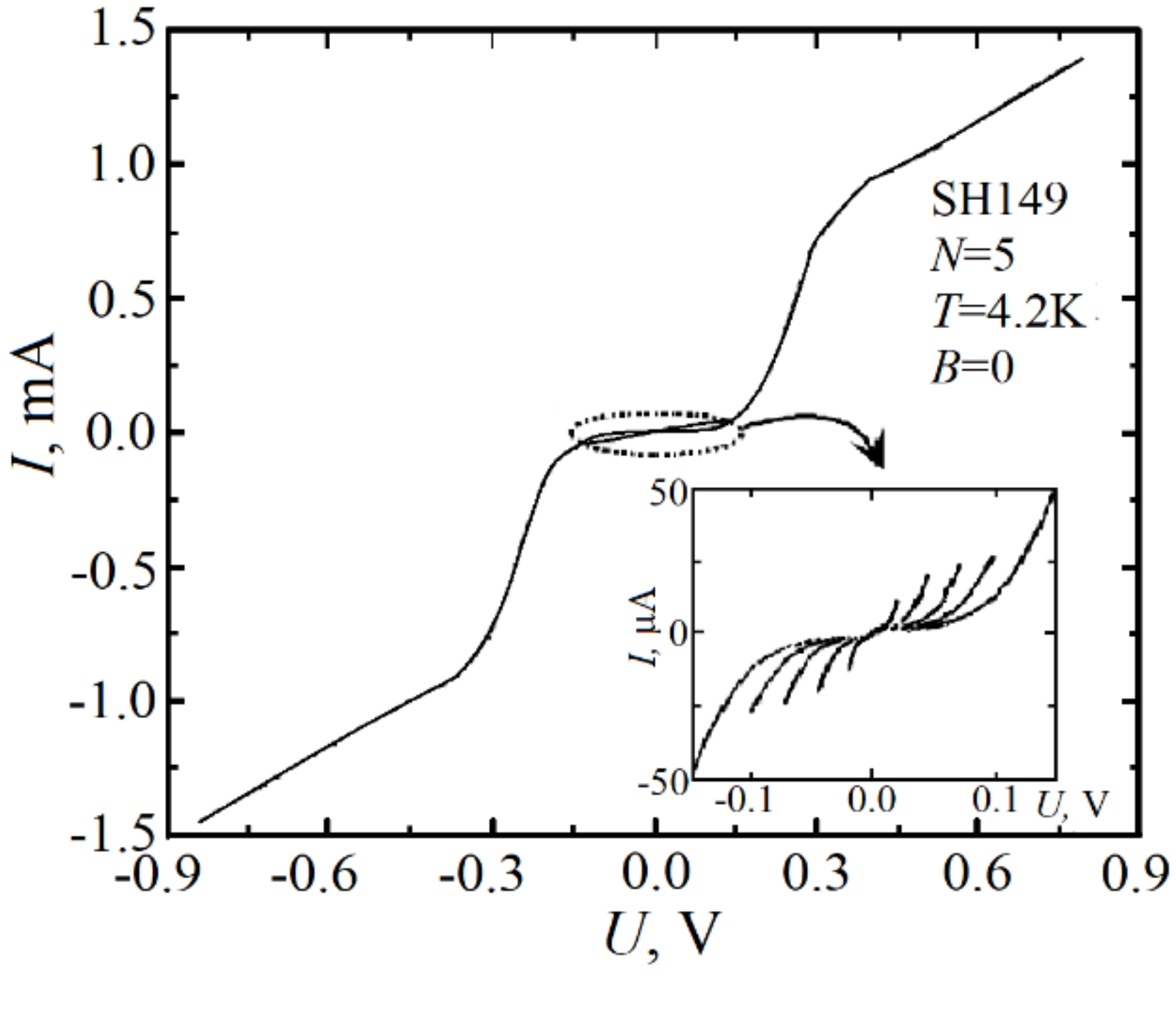

Fig. 3. Current-voltage characteristic of a BSSCO mesa structure measuring 1×1 µm$^2$, containing five junctions, at 4.2 K. The inset shows the same current-voltage characteristic on an enlarged scale (according to [11]).

The mesa structure shown in Fig. 3 contains N = 5 Josephson junctions; above the critical currents of these junctions, the CV characteristic has five branches in the resistive state, differing in the number of the resistive junction (see inset).

In the work of You et al. [15], a two-sided method for fabricating a stack of intrinsic Josephson junctions embedded in the bulk of very thin ($d \leq 100$ nm) $Bi_2Sr_2CaCu_2O_{8+x}$ single crystals was developed. The authors combined this method with precise etching control to create structures with a small (N = 0 - 6) number of intrinsic Josephson junctions, including the important case of a single junction. Figure 4 shows the CV characteristics of these structures with successively increasing numbers of intrinsic Josephson junctions. In addition to enabling superconducting gap tunneling spectroscopy, this technique is also important for potential applications such as SQUIDs and high-frequency mixers.

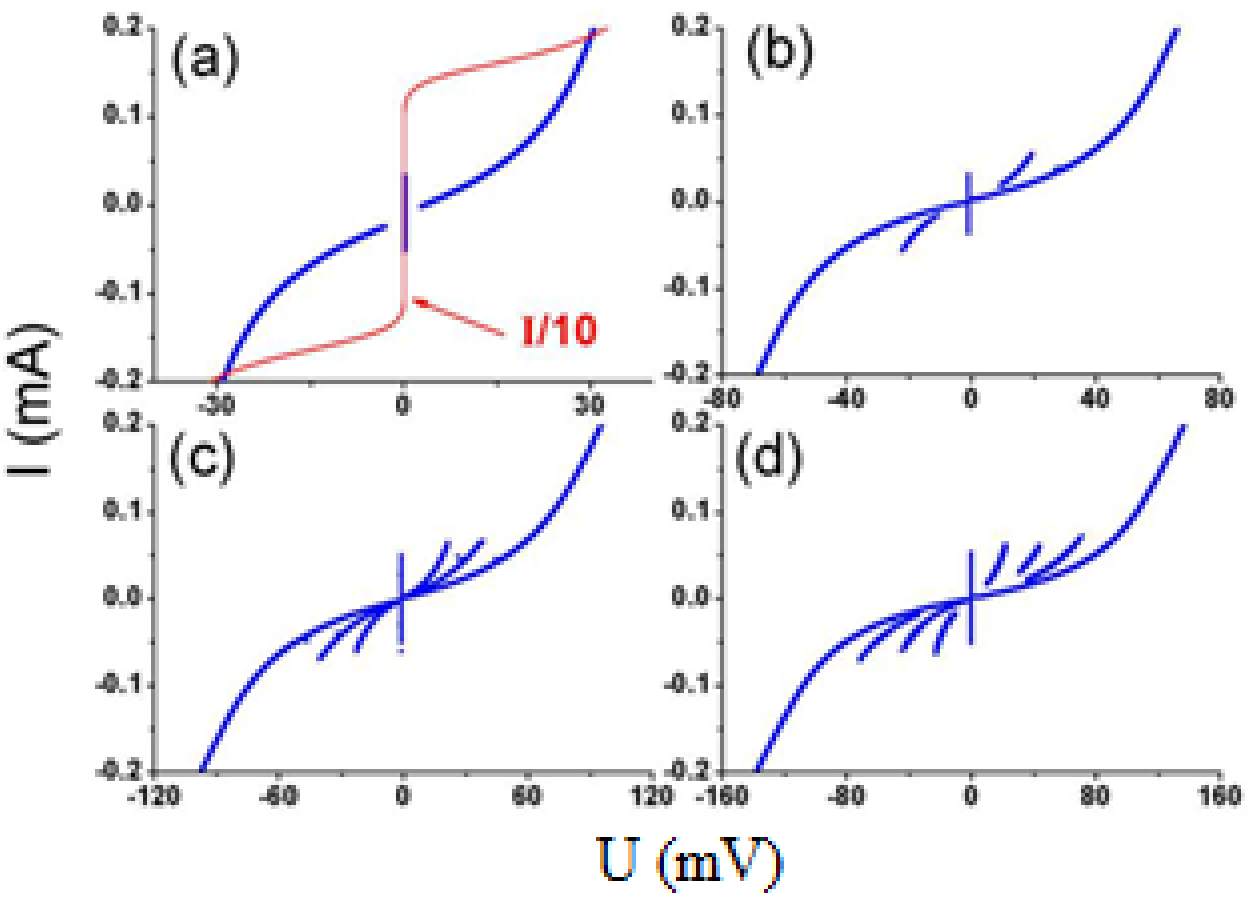

Fig. 4. The CV curves of the stack at 4.2 K with consecutively increasing number of junctions. The thin (red) curve in (a) shows the CV curve before the appearance of a single intrinsic Josephson junction (SIJJ). Note 10 times larger current scale for this curve (according to [15]).

While most of the experiments were devoted to the study of $Bi_2Sr_2CaCu_2O_8$ (BSCCO) systems, it was particularly important to study a compound other than BSCCO to provide the overall picture.

In [16], the intrinsic tunneling spectra of the La-doped $Bi_2Sr_{2-x}La_xCuO_{6+\delta}$(BSLCO) system with a single $CuO_2$ layer were presented for the first time. Despite a three-fold difference in the optimal critical superconducting temperatures for BSLCO and BSCCO (32 K and 95 K, respectively) and, consequently, different spectral energy scales, the authors discovered a clearly defined pattern of tunneling spectra for a stack of intrinsic Josephson junctions.

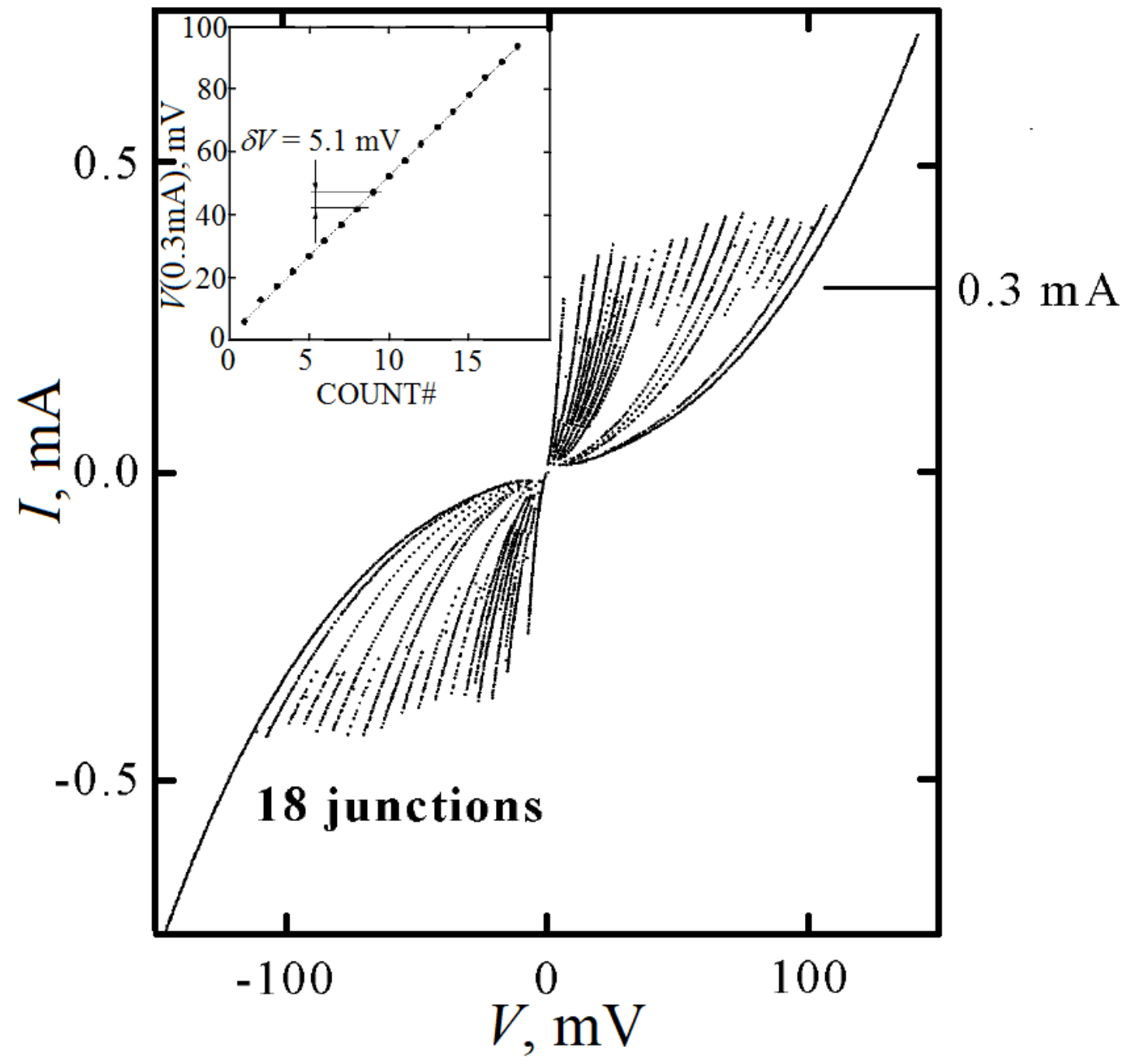

Fig. 5. CV characteristics for the BSLCO mesa structure with intrinsic tunnel junctions (N = 18) at $T = 4.2$ K. Inset: voltage measured at 0.3 mA for each calculated branch (N = 1...18). The linear $U$(N) dependence with equal intervals between points indicates the homogeneity of the Josephson junctions in the mesa structure (according to [16]).

Fig. 5 shows the current-voltage characteristics for an optimally doped BSLCO mesa structure with $T_c \approx 32$ K. At sufficiently low currents, there are N = 18 hysteresis branches, each of which corresponds to the sequential switching from the superconducting to the normal state of an increasing number of intrinsic tunnel junctions.

### *3.2 Current-voltage characteristics of electron-doped cuprate $Nd_{2-x}Ce_xCuO_4$*

Almost all the intrinsic Josephson junction (IJJ) characteristics have been observed in hole-doped cuprates and there are only isolated studies for electron-doped cuprates [19,20]. In an early study [19], Schlenga et al. observed the CV characteristics of the electron-doped high-$T_c$ superconductor $Pr_{2-x}Ce_xCuO_{4+\delta}$(PCCO) on a small platelet crystal. They observed the ac Josephson effect and several voltage steps in the CV characteristics, but not a multiple branch structure typical to IJJs.

The paper of Kawakami and Suzuki [20] reports the direct observation of the IJJ characteristics and their multiple branch structure for the electron-doped cuprate $Sm_{2-x}Ce_xCuO_{4-\delta}$(SCCO) by using a small mesa structure fabricated on a single crystal surface. This result indicates that the crystal structure of SCCO functions as superconductor-insulator-superconductor (SIS-type) IJJs like that of BSCCO. It was found that the CV characteristics in SCCO exhibit multiple resistive branches only when the lateral mesa width is less than a few micrometers.

We are interested in finding out the features of the manifestation of nonlinear dynamic properties of the intrinsic Josephson effect in the electron-doped high-temperature superconductor $Nd_{2-x}Ce_xCuO_4$(NCCO) with an optimal annealing. The $Nd_{2-x}Ce_xCuO_4$ single crystal has only one $CuO_2$ plane per unit cell and there are no apex oxygen atoms between adjacent conducting planes in the compounds with optimal annealing (see, for example, [1,28]**).** The coupling between cuprate planes is rather weak and a large anisotropy of the conductivity is observed indicating the quasi-two-dimensional nature of the electronic properties.

In our group, $Nd_{2-x}Ce_xCuO_{4+\delta}/SrTiO_3$ epitaxial films with $x = 0.145$ ($T_c = 15.7$ K) and 0.15 ($T_c = 23.5$ K) were synthesized by pulsed laser deposition [29,30]; the *c*-axis of the NdCeCuO lattice in these films was directed perpendicular to the $SrTiO_3$ substrate (standard orientation (001)) and along the long side of the SrTiO3 substrate (orientation $(1\bar{1}0)$).

We consider the $Nd_{2-x}Ce_xCuO_4$ crystal as a system of multiple quantum wells ($CuO_2$ layers) separated by doped Nd(Ce)O layers. The strong anisotropy of properties in the $Nd_{2-x}Ce_xCuO_4$ layered superconductor, with weak coupling between $CuO_2$ conductive layers, makes it possible to suggest that the NdCeCuO compound is a system with intrinsic Josephson junctions. This statement was proved by us as a result of the study of current-voltage characteristics for current flow in the direction of the *c*-axis.

In this Section, we present the current-voltage characteristics of $Nd_{2-x}Ce_xCuO_4/SrTiO_3$ epitaxial films with the *c*-axis of the NdCeCuO lattice directed along the sample for two Ce content ($x = 0.15$ and 0.145) at low temperatures $T$= (1.8–4.2) K in zero magnetic fields (see also [31]). The observed CV characteristics correspond to the standard superconductor-insulator–superconductor tunnel transitions and characterize the $Nd_{2-x}Ce_xCuO_4$ compound as a system with intrinsic Josephson junctions.

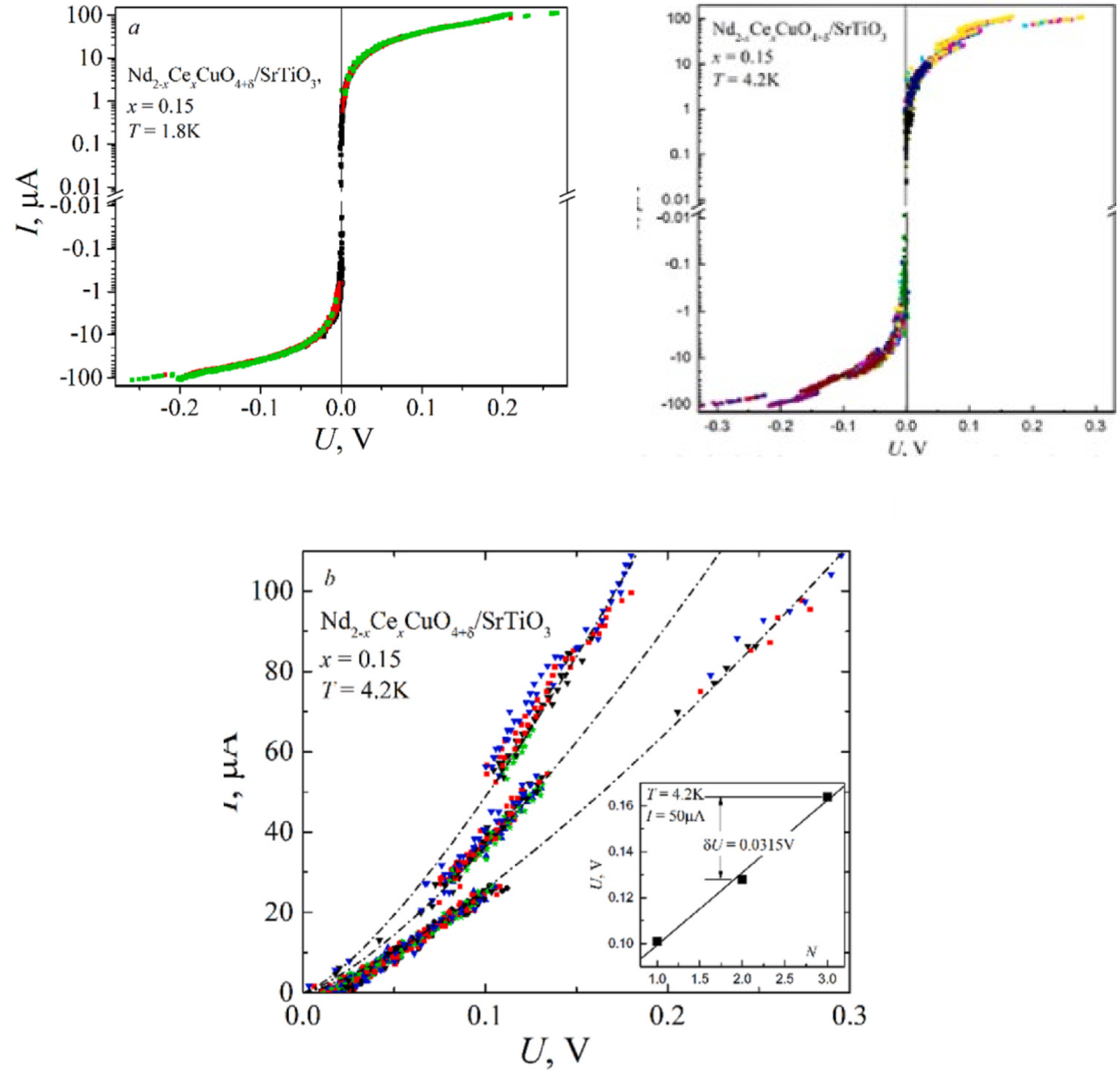

Fig. 6. (a) The common CV characteristics of $Nd_{2-x}Ce_xCuO_4/SrTiO_3$ film ($x$ = 0.15) at $T$ = 1.8 K and 4.2 K; (b) CV characteristic with three resistive branches for intrinsic tunnel junctions at $T$ = 4.2 K. Dashed lines correspond to the dependence $U \sim I^{0.75}$. Inset: the voltage measured at 50 μA for every counted branch (N = 1, 2,3) (after [31]).

Figure 6 illustrates the typical CV dependence for optimally annealed $Nd_{2-x}Ce_xCuO_4/SrTiO_3$ epitaxial films with $x$ = 0.15. The common CV characteristics at $T$ = 1.8 K and 4.2 K (Fig. 6a) demonstrate a zero-voltage supercurrent at $I < I_c$ with a low critical current value $I_c \cong 10\mu A$. In the CV characteristic for the current in the $c$ – direction the signs of intrinsic Josephson transitions are displayed in the region of $I > I_c$: N = 3 resistive branches are observed, separated by equal intervals $\delta U$ =31.5 mV at $I$ = 50 μA (see the inset on Fig. 6b). The linear dependence $U(N)$ indicates the uniformity of the junctions in the film. It should be noted the absence of hysteresis effects for the sample studied: the CV dependences are reversible for a given branch when the current increases and then decreases to zero.

From Fig. 6b, we see that the current rise is gradual for all branches with a high conductivity in the subgap regime. As discussed in Section 4, the functional dependence of $U$ on $I$ of this compound may indicate a nonmonotonic order parameter of $d$-wave symmetry inherent in $n$-type cuprates [32].

The CV characteristics for the optimally annealed $Nd_{2\text{-}x}Ce_xCuO_4/SrTiO_3$ epitaxial film ($x$ = 0.145) at $T$ = 1.8 K demonstrates multiple branches in the resistive state at $I > I_c$ (Fig. 7a, b). The CV curves are hysteretic that typical for Josephson tunnel junctions with large capacitive effects [24].

The precise CV measurements made it possible to visualize the jump from one resistive branch to other with an increase in current ($I = 0 \rightarrow I^*$) and back from one to other with a decrease in it ($I = I^* \rightarrow 0$) with $I^*$ = 20, 30, 45, 60, 70, 80 and 90 μA (see the top insert in Fig. 7b).

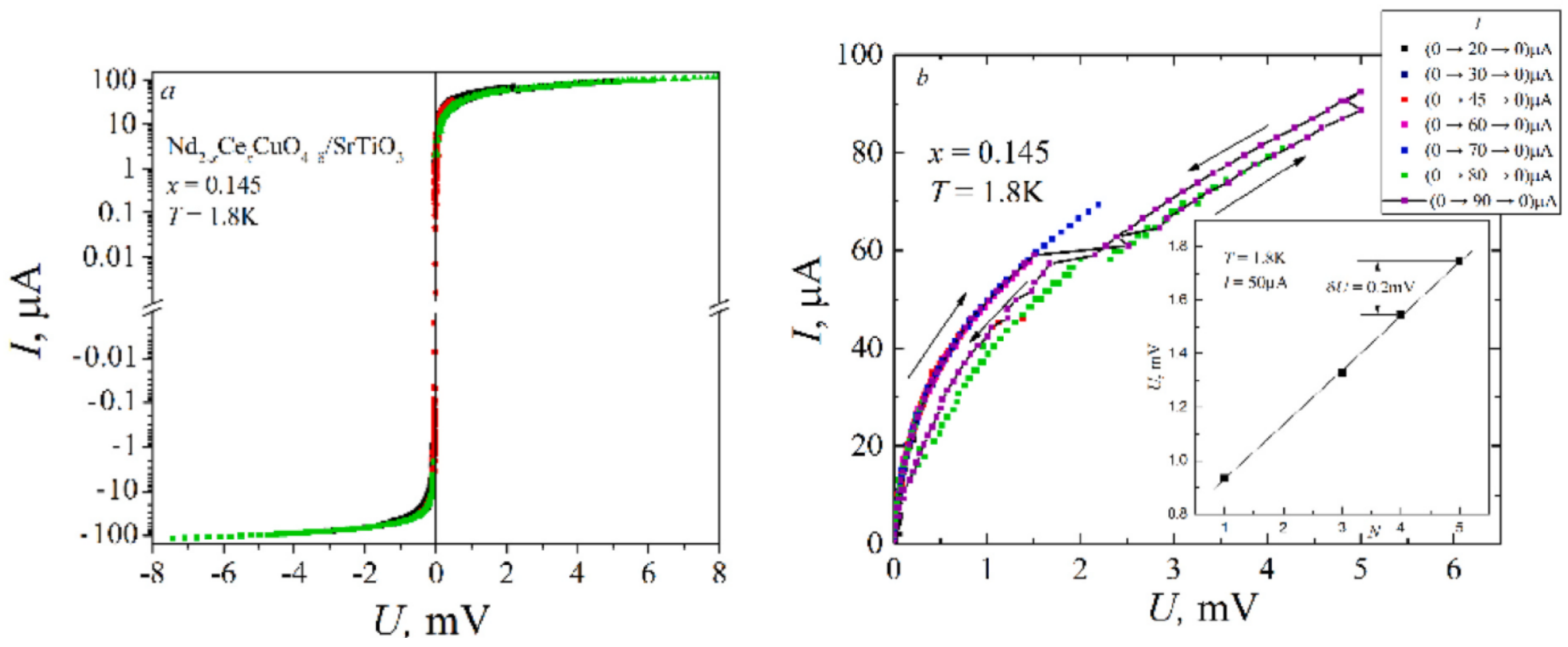

Fig. 7. (a) - The common CV characteristic for optimally annealed $Nd_{2\text{-}x}Ce_xCuO_4/SrTiO_3$ epitaxial film ($x$ = 0.145) at 1.8 K with zero-voltage supercurrent at $I < I_c \cong 40\mu A$.; (b) - hysteretic CV dependence and jumps from one resistive branch to other for this type of films at $T$ = 1.8 K. Inset: the voltage measured at 50 μA for every realizable branch (N = 1,3,4,5);

The resulting CV curves for $x$ = 0.145 clearly show the regularly spaced branches (with $\delta U$ = 0.2$mV$ at 50 $\mu A$) and it's possible to determine the numbers of observed junctions: $N$ = 1, 3, 4, 5 (see the bottom insert in Fig. 7b). The linear dependence $U(N)$ indicates the uniformity of the junctions in the film. This behavior corresponds to the standard superconductor-insulator–superconductor tunnel junction and characterizes the NdCeCuO compound as an array of resistively and capacitively shunted intrinsic Josephson junctions.

Several distinctive features should be noted in the behavior of the CV characteristics for the sample with $x$ = 0.145. The observation of hysteresis effects in Josephson tunnel junction is usually associated with a significant capacitive contribution, $I_d$, to the current in the resistive state at $I > I_c$ (see Eqs (3) and (4)), the measure of which is expressed in the value of the McCumber parameter $\beta_c$ [25,26].

Summarizing the results, we represent the parameters related to CV characteristics of the samples studied in Table 1.

Table 1. Parameters of $Nd_{2-x}Ce_xCuO_4$ films.

| $x$ | $\rho_c$ (mΩ·cm) | $\rho_c/\rho_{ab}$ | $R$ ($10^{-5}$Ω) | $C$ ($10^{-11}$F) | $I_c$ (μA) | $\beta_c$ |
|---|---|---|---|---|---|---|
| 0.15 | 2 | 100 | 4.3 | 4.2 | 10 | ~1 |
| 0.145 | 180 | 400 | 280 | 4.7 | 40 | ~25 |

Table 1 shows the following parameters: a degree of anisotropy, $\rho_c/\rho_{ab}$, the values of resistivity $\rho_c$ at $T = T_c^{\ onset}$, the resistance in the normal state $R = c\rho_c/S$ and the capacitance $C = \varepsilon S/c$ for a single junction, where $c = 60$ *nm* is the separation between $CuO_2$ layers and $S$ is the sample cross -sectional area, $\varepsilon = \varepsilon_0\varepsilon_r$, $\varepsilon_0$ is the *electric constant* ($\varepsilon_0 = 8.85 \times 10^{-12} Fm^{-1}$) and $\varepsilon_r$ is the dielectric constant of the material (we use for NCCO $\varepsilon_r \cong 10$ [4]). In addition, Table 1 shows the experimentally observed values of the critical current for Josephson junction systems, $I_c$, at 1.8 K, as well as the value of the McCumber parameter, $\beta_c$, calculated using formulas (4) and (2).

From the analysis of the data in Table 1 we firstly see the approximately twofold difference in the temperatures of SC transition, $T_c$, what is due to the different Ce content in the samples. And most importantly, there is a high difference in the values of the resistivities $\rho_c$ at $T = T_c^{\ onset}$, which further leads to a difference in the values of $R$, and, importantly, in the assessments of McCumber parameter $\beta_c$ (~ 25 times) in favor of $x = 0.145$ structure.

It is known [25,26] that the McCumber parameter is a measure for relative contributions of resistive and capacitive effects to the shunting of the Josephson junction at $I > I_c$ (see Eq. (3)). Strong differences in the value of the McCumber parameter ($\beta_c \sim 1$ for $x = 0.15$ and $\beta_c \sim 25$ for $x = 0.145$) lead to the implementation of different limiting cases of the RCSJ model in the samples studied (see numerical calculations in Ref. [26]).

For a case of $\beta_c \sim 1$, at $I > I_c$ the resistive component dominates in Eq. (3) ($I_n > I_d$), but for $\beta_c \sim 25$, on the contrary, the capacitive contribution plays the decisive role in the CV characteristic ($I_d \gg I_n$). Thus, the RSJ (resistively shunted junction) case is realized in $x = 0.15$ sample and the CSJ (capacitively shunted junction) case is settled in $x = 0.145$ sample, which leads to significantly different functional CV dependences in these samples.

In conclusion, the CV properties along the $c$ axis on $Nd_{2-x}Ce_xCuO_4/SrTiO_3$ epitaxial films with $x = 0.145$, 0.15 were investigated. For all the samples it has been established that the CV characteristics exhibit several resistive branches, which correspond to the resistive states of individual Josephson junctions.

The results confirm the idea of a tunneling mechanism between the $CuO_2$ layers (superconductor - insulator - superconductor junction) for the investigated $Nd_{2-x}Ce_xCuO_4$ compound.

## 4. Quasiparticle tunneling in *d*-wave superconductor-insulator-superconductor junctions

### *4.1 Coherent and noncoherent tunneling (theory)*

It is widely accepted that the order parameter of high-$T_c$ superconductors is primarily of *d*-wave symmetry, as supported by a large number of experimental results [33] (see, also. review [34] and references therein)

In superconducting tunnel junctions, the difference between *d*- and *s*-wave symmetry leads to a significant change in the CV characteristics. It is expected that a tunnel junction made of a *d*-wave superconductor exhibits a large subgap conductance in its CV curve due to the line nodes in the *d*-wave order parameter in a sharp contrast to the case of conventional *s*-wave tunnel junctions.

Indeed, this is supported by numerical calculations of CV characteristics for superconductor/insulator/superconductor (SIS) junctions with *d*-wave symmetry [35, 36]. The paper [36] presents a detailed discussion of the current-voltage characteristic of intrinsic Josephson junctions in $Bi_2Sr_2CaCu_2O_{8+\delta}$ and $Tl_2Ba_2Ca_2Cu_3O_{10+\delta}$. In these materials Josephson tunnel junctions are formed naturally between adjacent superconducting $CuO_2$ bilayers or trilayers and a typical sample consists of a stack of Josephson junctions.

Schlenga et al. [36] explicitly show that all junctions inside a given sample have identical tunneling characteristics and discuss the shape (general curvature) of the current-voltage characteristic in terms of a superconducting order parameter that has a predominant $dx^2 - y^2$ symmetry.

In the resistively shunted junction (RSJ) model [24, 25] the current through a Josephson junction is given as the sum of the supercurrent $Is$, the quasiparticle current $I_n$, and the displacement current $I_d$ (see Eq. (3)). In most cases the resistive branch of the CV characteristic is completely dominated by the quasiparticle current and the other two channels cause only minor deviations from the pure quasiparticle characteristic.

Therefore, the explanation of the general curvature is reduced to that of the quasiparticle tunneling characteristic. At low temperatures the tunneling characteristic of the intrinsic Josephson junctions in BSCCO as well as in TBCCO shows a high conductivity in the subgap regime ($V \ll 2\Delta/e \approx 50mV$).

This contrasts with a pure BCS-like density of states, which leads to a steep current rise in the tunneling characteristic at a voltage $2\Delta/e$. To compare experimental results to theory, the tunneling quasiparticle current $I(U)$ in the $c$-axis direction was calculated in [36] using the expression:

$$I(U) = \frac{1}{eR_n}\int_{-\infty}^{\infty} N(E)N(E+eU)[f(E)-f(E+eU)]dE. \tag{5}$$

Here, $R_n$ is the normal state resistance, $f(E)$ the distribution function, and $N(E)$ the normalized density of states of the superconductors.

Both experiments probing the quasiparticle excitation spectrum as well as phase sensitive experiments provide evidence that the order parameter in high-$T_c$ superconductors has $d_{x^2-y^2}$ symmetry [33], [34]. Thus, for the normalized density of states $N(E)$ Schlenga et al. used the simplified expression for $d_{x^2-y^2}$ -wave superconductors, [37]:

$$N(E) = \frac{1}{2\pi}\int_0^{2\pi} N(E.\varphi)d\varphi, \tag{6a}$$

$$N(E.\varphi) = Re\left[\frac{E}{\sqrt{E^2-\Delta_0^2 cos^2 2\varphi}}\right], \tag{6b}$$

where $\varphi$ is the in-plane angle in $\boldsymbol{k}$ space, $N(E.\varphi)$ is the partial density of states and $\Delta_0$ is the maximal value of the energy gap parameter.

Yamada and Suzuki [38] paid attention to the fact that the calculations of Schlenga et al. [36] lack an important factor, specifically, the coherence of tunneling, or the conservation of the transverse momentum. Their paper presents numerical calculations of the coherent quasiparticle tunneling in $d$-wave SIS junctions for comparison with experimental results. It is shown that the coherent tunneling constraint causes an essential change in the CV characteristics.

In the coherent tunneling, the momentum perpendicular to the tunnel direction $\boldsymbol{k}^{\perp}$ is conserved so that $\boldsymbol{k}_L^{\perp} = \boldsymbol{k}_R^{\perp}$ or $k_L^{\perp} = k_R^{\perp}$ and $\varphi_L = \varphi_R$ ($\varphi$ is the in-plane angle in $k$ space and suffixes $L$, $R$ denote left and right, respectively). Then the tunneling matrix element may be rewritten as

$$|t(\boldsymbol{k}_L^{\perp}.\boldsymbol{k}_R^{\perp})|^2 = (2\pi)^2|t|^2\delta(k_L - k_R)\delta(\varphi_L - \varphi_R). \tag{7}$$

After transfiguration of the original formulas, the following expression for the coherent quasiparticle current was obtained:

$$I(U) = \frac{1}{2\pi eR_n}\int_0^{2\pi} d\varphi \int_{-\infty}^{\infty} dE\, N(E,\varphi)N(E+eU,\varphi)[f(E)-f(E+eU)], \tag{8}$$

where, in $d$-wave superconductors, $N(E.\varphi)$ is expressed by Eq. (6b).

From numerical calculations it was found that in coherent tunneling the differential conductance *dI/dU* exhibits a very sharp peak at the superconducting gap voltage in a sharp contrast to the case of incoherent tunneling, where the *dI/dU* peak is very broad (see Figs 1 and 2 in [38] with *dI/dU-U* curves for coherent and incoherent cases, respectively). The sharp *dI/dU* peak is in good agreement with experimental results for intrinsic Josephson junctions of $Bi_2Sr_2CaCu_2O_{8+\delta}$ [39].

Another difference in the tunneling characteristics is seen in the subgap conductance $\sigma(U) \equiv dI/dU$ for $U < 2\Delta/e$. In the coherent tunneling case, $\sigma(U)$ is almost linear in $U$ [38], while in the incoherent tunneling case, $\sigma(U)$ is nearly proportional to $U^2$ for $U < 2\Delta/e$ [36].

Yamada and Suzuki [38] also considered the case of some deviation from the coherency where a small shift in φ is involved in the process of tunneling due to imperfections at the junction interfaces which scatter quasiparticles. The tunneling probability becomes proportional to the Gaussian distribution:

$$|t(\varphi_L, \varphi_R)|^2 = 2\pi \frac{|t|^2}{\sqrt{2\pi}S} exp\left\{-\frac{\theta^2}{2S^2}\right\} \equiv 2\pi h(\theta), \qquad (9)$$

where $\theta = \varphi_L - \varphi_R$ , and $S$ is the measure of φ deviation. Large $S$ reflects a large number of scattering centers at the interfaces. Then we have:

$$I(U) = \frac{1}{2\pi e R_n} \int_0^{2\pi} d\varphi \int_{-\infty}^{\infty} d\theta\, h(\theta) \int_{-\infty}^{\infty} dE\, N(E, \varphi) N(E + eU, \varphi + \theta) \times$$
$$\times [f(E) - f(E + eU)] \qquad (10)$$

In [38] it is shown that CV and *dI/dU-U* curves smoothly change with increasing *S* from the coherent tunneling curves to the incoherent ones and for *S*=30° and larger, the tunneling characteristics are almost identical to those of incoherent tunneling. So, the tunneling characteristics are very sensitive to the fraction of incoherent tunneling.

### *4.2 Scaling behavior of intrinsic current-voltage characteristics (experiment)*

Thus, within the confines of standard tunneling formalism, quasiparticle current for SIS tunnel junction with $d_{x^2-y^2}$-wave density of states for the superconductors is numerically calculated both for coherent [38] and incoherent [36, 38] tunneling cases. The coherency (or incoherency) of tunneling process is defined by the conservation (or nonconservation) of the transverse momentum. Calculations indicate that for $U < 2\Delta/e$ the subgap conductance $\sigma(U) \equiv dI/$ $dU$ is almost linear in $V$ in the coherent tunneling case [38], while $\sigma(U)$ is nearly proportional to $V^2$ for incoherent tunneling [36, 38].

Then, theoretically, for a general curvature of a branch in the current voltage characteristic of SIS tunnel junction with a superconducting order parameter that has a predominant $d_{x^2-y^2}$ symmetry we should have $U \sim I^\gamma$, where $\gamma \approx 1/2$ for coherent tunneling and $\gamma \approx 1/3$ for incoherent case.

*4.2.1 Hole-doped cuprates*

In [36, 38] a numerical fit of the calculated CV dependences to selected experimental curves was carried out. To reveal the general curvature of CV dependences for branches in the current-voltage characteristic we presented a large set of relevant experimental data for hole-doped cuprate HTSCs on a double-logarithmic scale (see Fig.8 and Table 2).

We used data of works [10], [36] for Tl - systems and of works [11], [40] for Bi - systems. It was found that for such a set of hole-doped samples the CV curves for individual branches in the current-voltage characteristic fits the power-law dependence $U \sim I^\gamma$ with $\gamma = (0.29–0.38)$ in good agreement with theoretical calculations ($\gamma \approx 1/3$) for the case of incoherent tunneling. For high-quality $Bi_2(Sr_{2-x}La_x)CuO_6$ mesa-structures [16] the treatment of CV dependences for an array of resistive branches leads to the value $\gamma = 0.48$, which may correspond to the case of coherent tunneling processes with $\gamma \approx 1/2$ [38].

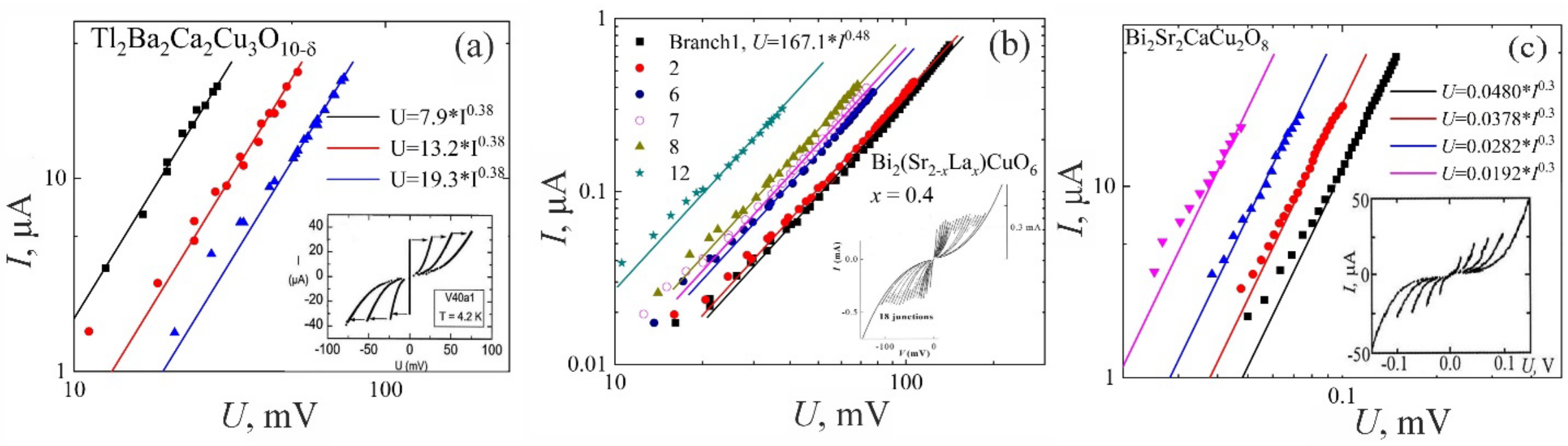

Fig.8. Scaling behavior of the intrinsic CV characteristics for different branches of the compounds: (a) $Tl_2Ba_2Ca_2Cu_3O_{10+\delta}$, inset – initial *I(U)* dependencies [36]; (b) $Bi_2(Sr_{2-x}La_x)CuO_6$, inset – initial *I(U)* dependencies [16]; (c) $Bi_2Sr_2CaCu_2O_8$, inset – initial *I(U)* dependencies [11].

Table 2. Dependence index $U \sim I^\gamma$ for various hole-doped HTSC compounds

| Compound | γ | Refs |
|---|---|---|
| $Bi_2(Sr_{2-x}La_x)CuO_6$ | 0.48 | [16] |
| $Tl_2Ba_2Ca_2Cu_3O_{10}$ | 0.38 | [10], [36] |
| $Bi_{2.43}Sr_{1.85}Ca_{0.90}Cu_{1.82}O_x$ | 0.36 | [40] |
| $Bi_2Sr_2CaCu_2O_8$ | 0.30 | [11] |

*4.2.2 Electron-doped cuprates*

In contrast with hole doped cuprates with simple *monotonic* $d_{x^2-y^2}$ symmetry, the interpretation of microwave measurements along with angle- resolved photoemission spectroscopy (ARPES), phase sensitive scanning SQUID microscope experiments and low energy electronic Raman scattering studies for the electron doped cuprates is consistent with *nonmonotonic* form of the *d*-wave SC order parameter (see, for example, Fig. 9 and extensive references in [32]). μ

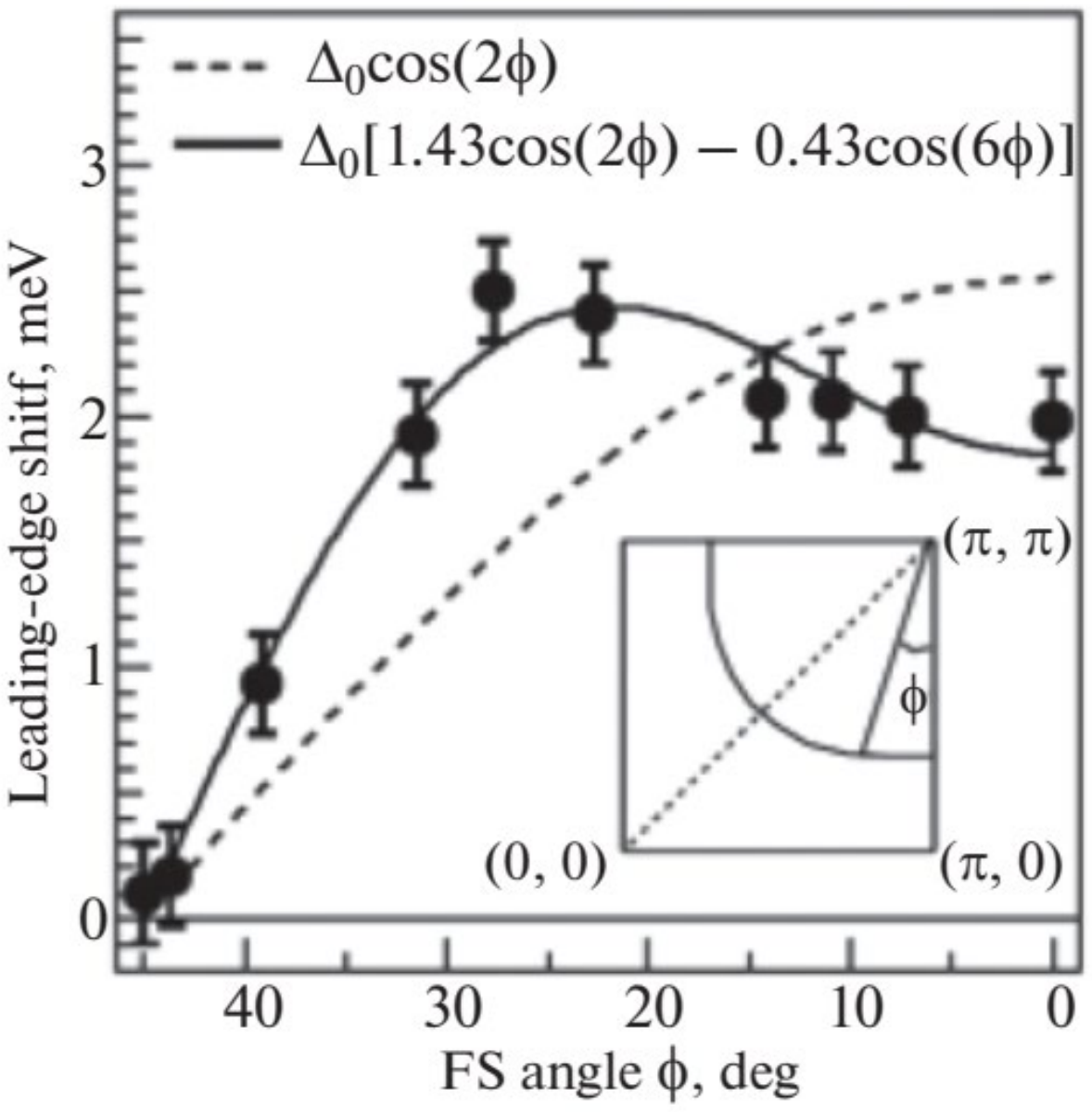

**Fig. 9.** Superconducting gap of PLCCO from ARPES. Leading-edge shift determined as a function of position (angle) on the Fermi surface showing that it fits a nonmonotonic *d*-wave symmetry (after [41]).

It has been first argued by Yuan et al. [42] that the experimentally observed nonmonotonic gap in electron doped cuprates at optimal doping is the lowest quasiparticle excitation energy in the *coexisting* antiferromagnetic (AFM) and superconducting (SC) state. When AFM order coexists with the SC order, the resulting quasiparticle excitation can be gapped by both orders and behave to be nonmonotonic *d*-wave, even though the SC gap itself could have a typical monotonic $d_{x^2-y^2}$ symmetry (see references in [43]).

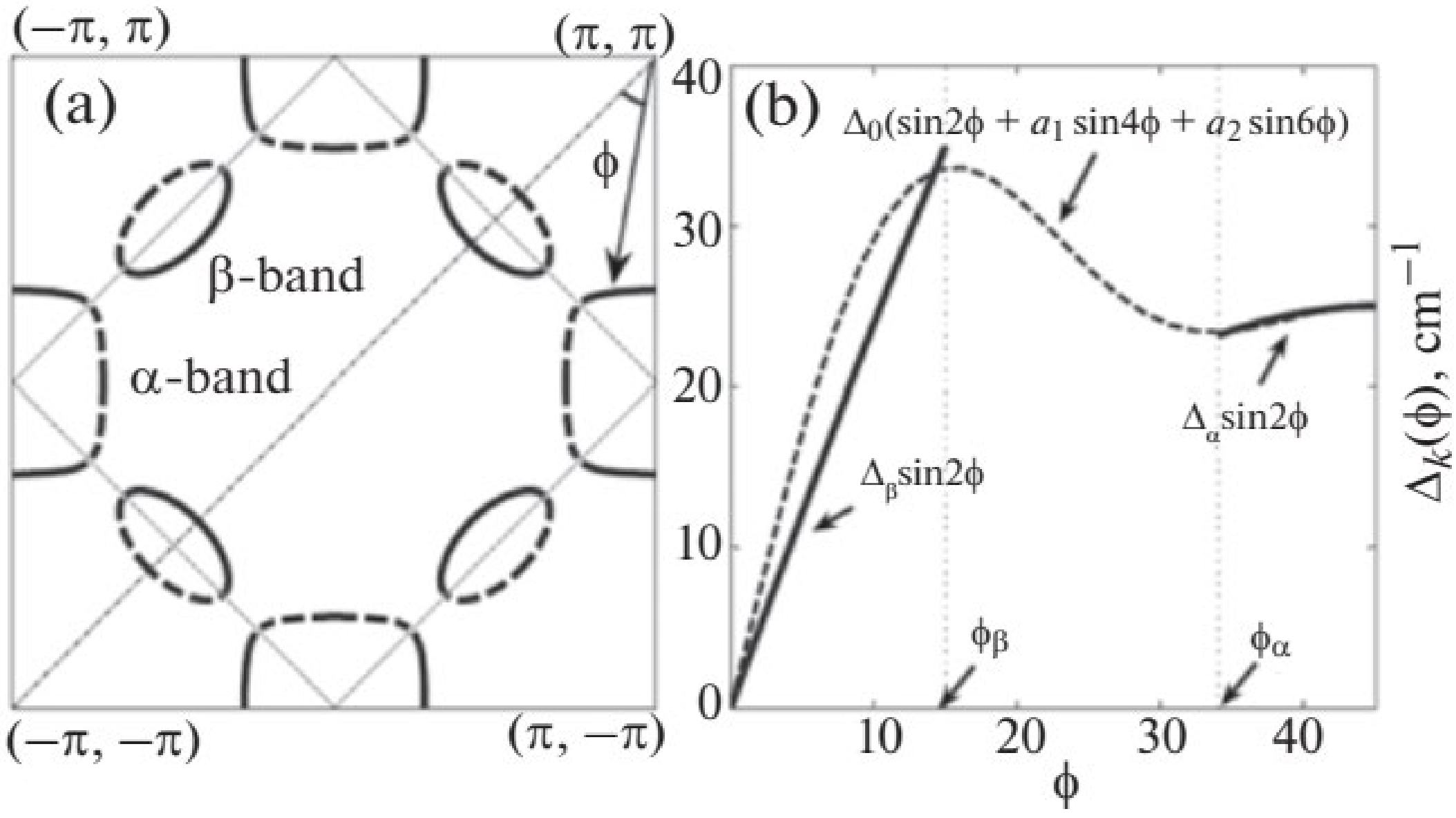

**Fig. 10.** (a) Separate α- and $\beta$-band FSs of the electron doped cuprates with an AFM order. (b) Gap function $\Delta_k(\phi)$ with ϕ measured along the FS: (dashed curve) a nonmonotonic *d*-wave $\Delta_k$; (solid lines) the piecewise monotonic $d_{x^2-y^2}$-wave $\Delta_k$ (after [43]).

A pairing model considering AFM spin fluctuation (*Two-Band/Two-Gap Model)* is proposed for the electron doped cuprate superconductors by Liu and Wu [43]. It suggests that, similar to the hole doped side, the superconducting gap function is monotonic $d_{x^2-y^2}$ -wave and explains why the observed gap function has a nonmonotonic *d*-wave behavior when an AFM order is taken into account.

The existence of the two doping-dependent FS in electron doped cuprates revealed by ARPES is well explained in terms of the band-folding effect associated with an AFM order which splits the band into upper (electron-like) and lower (hole-like) branches. Two types of FS parts in superconducting state of electron doped cuprates presented in the figures from [43] (see Fig. 10).

Figure 10a shows the FSs for a typical optimally doped sample. The FS continuum splits into electron-like band (α-band) that crossed by the Fermi level in the antinodal region and hole-like band (β-band) that crossed by the Fermi level in the nodal region. Figure 10b shows an example of piecewise monotonic $d_{x^2-y^2}$-wave $\Delta_k$, compared with a nonmonotonic one.

We emphasize that the model of Liu and Wu suggests the *monotonic* $d_{x^2-y^2}$ -wave SC gap function for each piece of FS (both for electron-like and hole-like parts) and explains that the observed gap function has a *nonmonotonic* *d*-wave behavior due to coexistence of superconductivity with the AFM order.

Onwards in this section, we present our analysis of data on subgap current for intrinsic Josephson tunneling in electron- doped cuprate superconductors $Sm_{2-x}Ce_xCuO_4$ [20], $Pr_{2-x}Ce_xCuO_4$ [19] and $Nd_{2-x}Ce_xCuO_4$ [31]. Among the $CuO_2$ layered compounds the *n*-doped materials are of special interest not only by their different species of carriers but also by the fact that there are only single $CuO_2$ layers.

Kawakami and Suzuki [20] investigated the current-voltage characteristics of the intrinsic Josephson junctions in the electron-doped high-$T$c superconductor $Sm_{2-x}Ce_xCuO_{4-\delta}$ by using a small mesa structure fabricated on a single crystal surface. Figure 11 shows the CV characteristics for $10 \times 10\ \mu m^2$ mesa with the three-terminal (3T) type electrode configuration found in [20] and the CV dependences we constructed on a double logarithmic scale for subgap regime in the Sm – structure. Rather well-expressed scaling behavior, $U{\sim}I^{\gamma}$, with $\gamma \approx 0.5$ at $U < 30mV$ and $\gamma = 0.7$ for larger values of $U$ is observed.

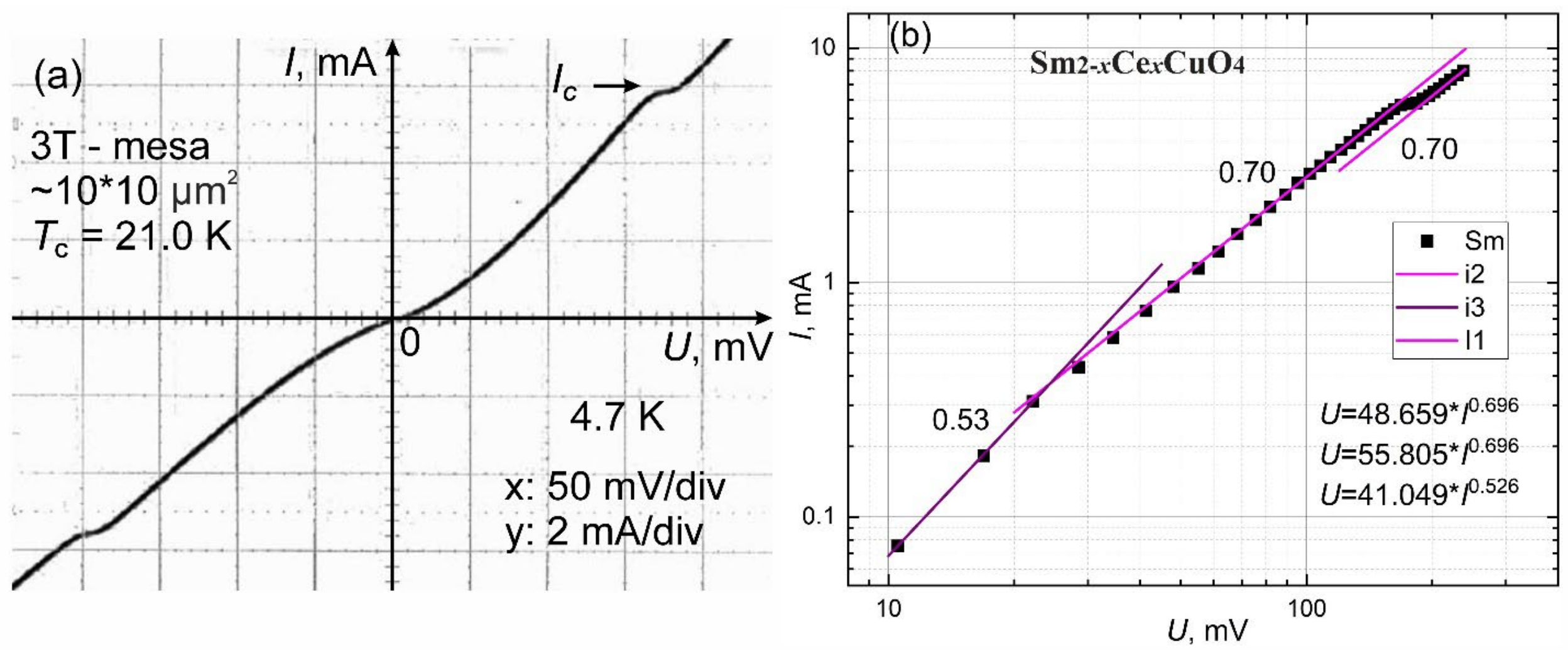

**Fig. 11.** (a) Experimental dependences $I(U)$ for the electron-doped compound $Sm_{2-x}Ce_xCuO_4$ (after [20]); (b) scaling behavior, $U{\sim}I^{\gamma}$, of intrinsic CV characteristics for $Sm_{2-x}Ce_xCuO_{4-}$ single crystal at $T = 4.7$ K.

We obtained interesting results from the analysis of data on CV characteristics in the Pr-systems [19]. Fig. 12(a) shows experimental data on the $c$-axis transport for optimally doped $Pr_{2-x}Ce_xCuO_{4+\delta}$ single crystals (with $T_c$ ranged from 16 K to 18 K) from the work of Schlenga et al. [19]. Upon increasing the current bias there are several voltage jumps in the CV characteristic due to junctions with relatively small critical currents. They are followed by a jump of about 48mV where the main part of the intrinsic junctions becomes resistive. The CV characteristic shows only a small hysteresis. The microwave emission in the X band frequency range was detected.

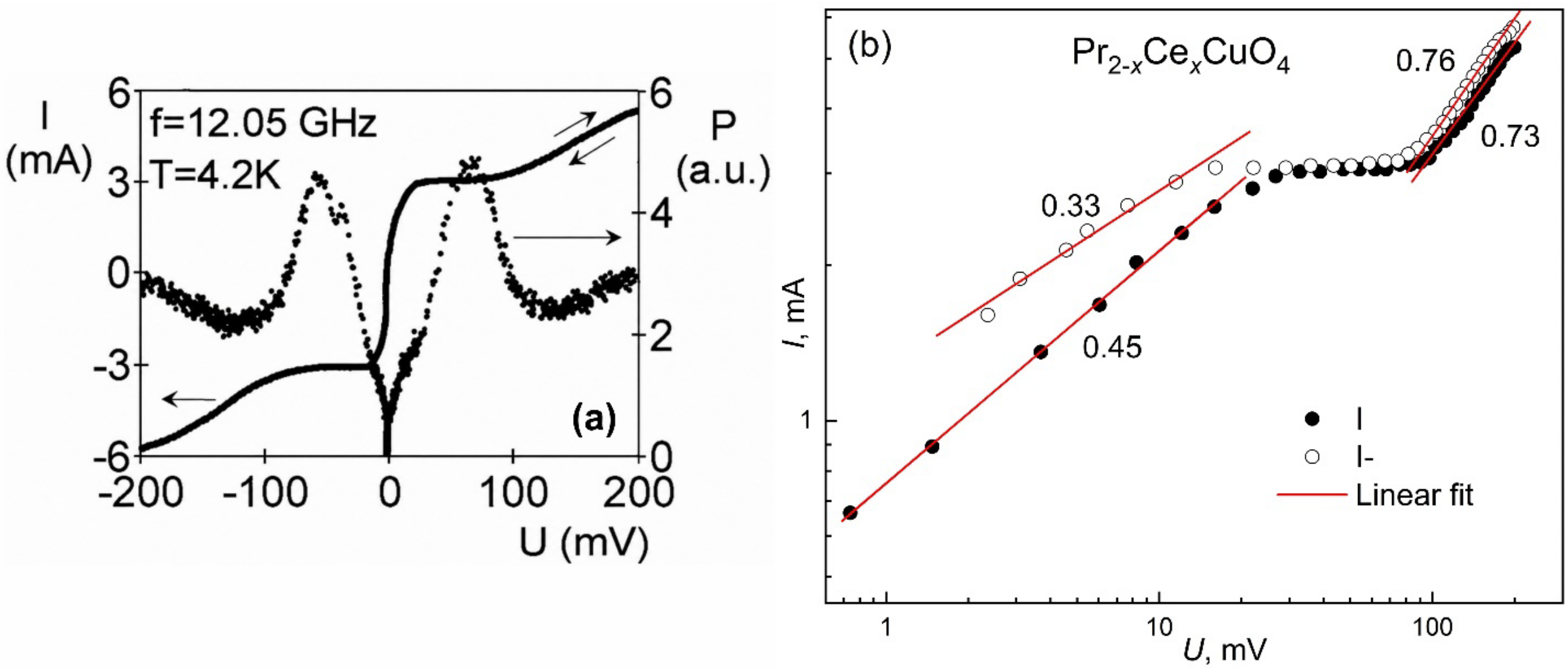

**Fig.12.** (a) CV characteristic with voltage biasing (left scale) and detected microwave power (f=12.05GHz, right scale) of an $Pr_{1.85}Ce_{0.15}CuO_{4+\delta}$ single crystal. The voltage at the peak maximum corresponds to emission of about 2700 junctions (after [19]); (b) scaling behavior, $U{\sim}I^{\gamma}$, of intrinsic CV characteristics for optimally doped $Pr_{2\text{-}x}Ce_xCuO_{4+\delta}$ single crystal at $T = 4.2$ K.

On Fig.12(b), our analysis of these data ($I$ *vs* $U$ log-log plot) presented. Here we see two well-defined regions of power law dependence $U{\sim}I^{\gamma}$ with $\gamma = 0.45\ (0.33)$ at $U < 20mV$ and $\gamma = 0.73\ (0.76)$ at $U > 100mV$ for $I^{+}(\,I^{-})$, respectively. We think that such a behavior both in the SCCO and in PCCO crystals is due to the symmetry features of the SC order parameter for n-type cuprates (see above).

Based on the two-band/two-gap model for the spectrum of electron-doped cuprates (see Fig. 10), we can assume that in the $U{\sim}I^{\gamma}$dependences at small $U$ we observe tunneling processes for a single branch of the gap function $\Delta_k(\phi)$ with $d_{x^2-y^2}$-symmetry (the results for $\gamma$ are within the limits $1/3 < \gamma < 1/2$ as in the theory [36. 38] and for hole-doped cuprates, see Table 2).

At the same time, the dependences of $U(I)$ at larger $U$ reflect a more complete picture of the non-monotonic $d$-symmetry for $\Delta_k(\phi)$ when both gaps $\Delta_\alpha$ and $\Delta_\beta$ are involved. It is logical to associate the much larger values of $\gamma$ = (0.7 - 0.76) we obtain for SCCO, PCCO and NCCO (see below) with a specific type of $d$-symmetry of the order parameter, inherent for electron-doped cuprates.

For the current-voltage characteristics studied in our work [31] for epitaxial film of electron-doped cuprate $Nd_{2\text{-}x}Ce_xCuO_4$ ($x$ = 0.15) with conductivity along the $c$-axis, we see the same trend. In Fig.13, the CV dependence for the film with $x$ = 0.15 is presented on a double logarithmic scale. We see that when $U > \sim 0.05$ V experimental data for all three branches fit on straight lines corresponding to the dependence $U \sim I^{0.75}$. Dependencies of this type are given in Fig. 6 b by the dashed lines.

The spread of experimental points at very low currents ($I <\sim 5\ \mu$A) and voltage ($U <\sim 0.05$ *V*) in Fig.13. is due to random fluctuations in the values of the critical current $I_c$ when the transition to a resistive state is turned on. Nevertheless, in this region, one can identify tendencies towards dependence $U \sim I^{\gamma}$ with $\gamma \approx 0.35$ (see the red line in the main part of the figure, as well as the set of straight lines in the inset).

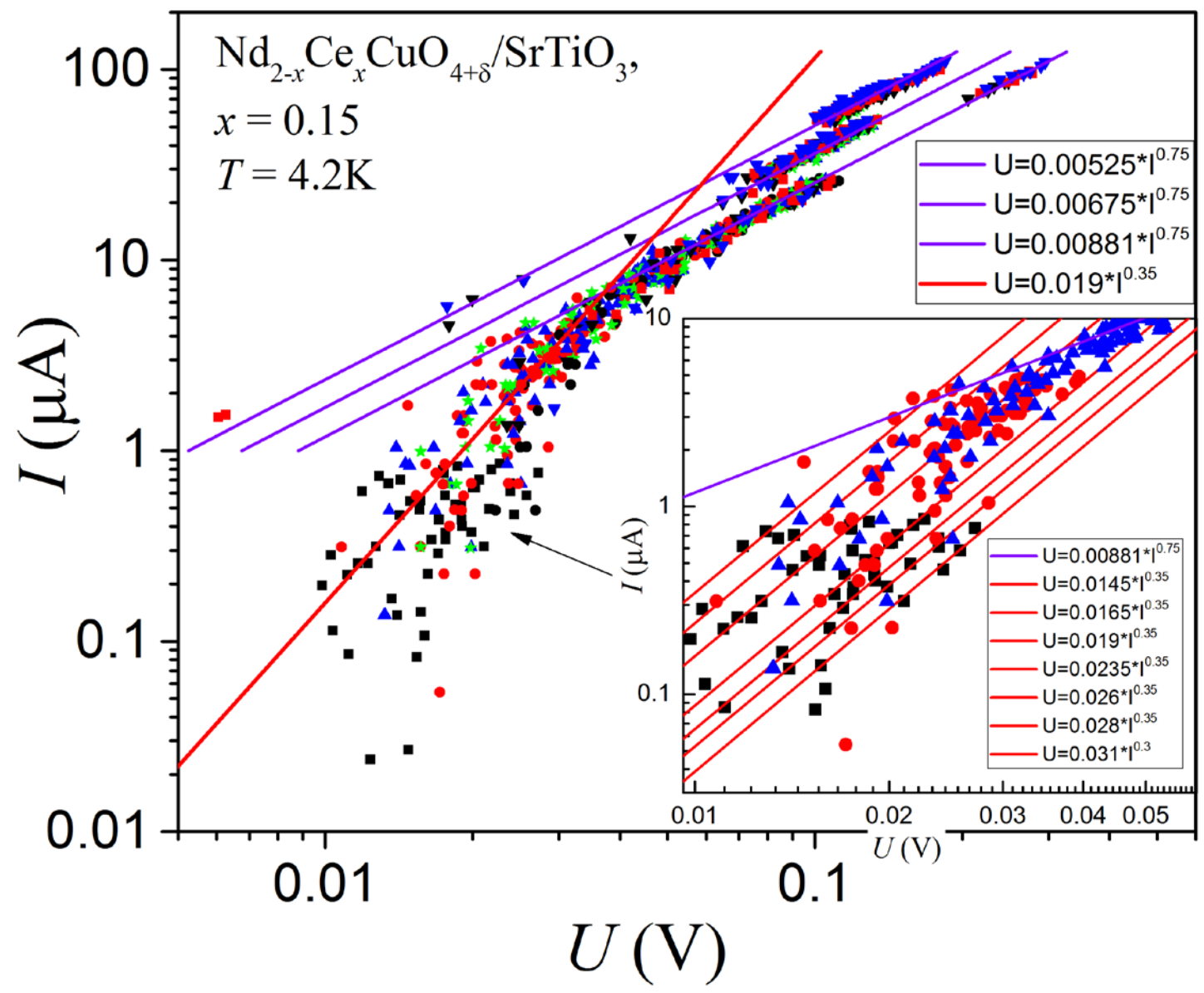

**Fig.13.** Scaling behavior of intrinsic CV characteristics for different branches of $Nd_{2-x}Ce_xCuO_4$/$SrTiO_3$ epitaxial film ($x = 0.15$) at $T = 4.2$ K (according to data from [31]). Inset: *I vs U* dependencies of the film at $U < 0.05$ V.

To summarize, in Table 3 we present the values of parameter $\gamma$ for the discussed electron-doped HTSC compounds.

Table 3. Dependence index $U \sim I^{\gamma}$ for various electron-doped HTSC compounds

| Compound | γ low U | γ high U | Refs |
|---|---|---|---|
| $Sm_{2-x}Ce_xCuO_4$ | 0.52 | 0.7 | [20] |
| $Pr_{2-x}Ce_xCuO_4$ | 0.45 ($I^+$) | 0.73 ($I^+$) | [19] |
| | $0.33(I^{-})$ | $0.76(I^{-})$ | |
| $Nd_{2-x}Ce_xCuO_4$ | 0.35 | 0.75 | [31] |

## 5. Conclusions

In the first part of the review, we examined the observation of multi-branch current-voltage characteristics for current in the *c*-direction both for the hole-doped cuprate SCs and for our epitaxial films of electron-doped SC $Nd_{2-x}Ce_xCuO_4$. This phenomenon indicates that the layered SCs under consideration may be regarded as the stack of intrinsic Josephson junctions with the layers of $CuO_2$ being the SC plates, and the buffer layers being the insulator spacers.

It's found that the tunneling characteristics of the intrinsic Josephson junctions in the optimally doped $p$- and $n$ - type cuprates at low temperatures show a high conductivity in the subgap regime $U < 2\Delta/e$. This indicates the predominantly $d$-wave symmetry of the SC order parameter in optimally doped cuprates as an existence of the nodes in the gap for the $d$-wave symmetry leads to a large subgap tunnel current in contrast to the $s$-wave case.

In the second part of the review, we carried out a quantitative analysis of the subgap tunneling conduction by plotting the graphs $I(U)$ on a double logarithmic scale. The analysis carried out confirms the theoretically predicted power-law dependences, $U \sim I^{\gamma}$, for the $d_{x^2-y^2}-$ wave pairing symmetry (with 0.33< $\gamma$< 0.5 from incoherent to coherent tunneling) in the hole-doped Bi- and Tl- systems.

Our analysis showed a peculiar result for the electron-doped SCs, for which the theory (and the phase-sensitive experiments) suggests a non-monotonic $d$-wave symmetry of the SC order parameter within the framework of the two-band/two-gap model.

For the Sm- and Pr- systems, as well as for our Nd-films with $x$=0.15, we see the manifestation of tunneling through a single gap ($U \sim I^{\gamma}$ with $\gamma$ = 0.33–0.5) at small $U$, but a more complex picture, most likely corresponding to the manifestation of the two-gap nature of the SC spectrum, turns out at large $U$ ($U \sim I^{\gamma}$ with the higher values of $\gamma$ = 0.7–0.75).

In general, it can be said that the observed features of the CV characteristics for the current in the *c*-direction in layered hole- and electron-doped cuprates are due to the manifestation of the $d$-wave SC pairing type in tunneling processes for the stacks of intrinsic Josephson junctions.

**Acknowledgments**

The work was carried out within the framework of the state assignment of the Ministry of Science and Higher Education of the Russian Federation for the IMP UB RAS.